\documentclass[aps,prb,reprint,superscriptaddress,nofootinbib]{revtex4-2}

\usepackage{amsmath} 
\usepackage{graphicx} 
\usepackage{float}
\usepackage{helvet}
\usepackage{makecell}
\usepackage{multirow}
\usepackage{siunitx}
\usepackage{float}
\usepackage{tablefootnote}

\usepackage{soul}
\usepackage[usenames, dvipsnames]{xcolor}

\usepackage[version=4]{mhchem}

\begin{document}

\title{Atomic cluster expansion potential for the Si--H system}

\author{Louise A. M. Rosset}
\affiliation{Inorganic Chemistry Laboratory, Department of Chemistry, University of Oxford, Oxford OX1 3QR, UK}

\author{Volker L. Deringer}
\email{volker.deringer@chem.ox.ac.uk}
\affiliation{Inorganic Chemistry Laboratory, Department of Chemistry, University of Oxford, Oxford OX1 3QR, UK}

\begin{abstract}
The silicon--hydrogen system is of key interest for solar-cell devices, including both crystalline and amorphous modifications. Elemental amorphous Si is now well understood, but the atomic-scale effects of hydrogenating the silicon matrix remain to be fully explored. Here, we present a machine-learned interatomic potential model based on the atomic cluster expansion (ACE) framework that can describe a wide range of Si--H phases, from crystalline and amorphous bulk structures to surfaces and molecules. We perform numerical and physical validation across a range of hydrogen concentrations and compare our results to experimental findings. Our work constitutes an advancement toward the exploration of large structural models of $a$-Si:H at realistic device scales.
\end{abstract}

\maketitle

\section{Introduction}

Amorphous silicon (a-Si) is a prototypical disordered material which continues to be of interest for a wide range of applications, from interferometer mirror coatings \cite{Birney-18-11} to novel solar-cell heterojunction technologies \cite{lin_silicon_2023, Fischer-23-03}.
In these devices, the amorphous matrix often contains a varied concentration of hydrogen, from 5 to 20 at.-\% H \cite{street_hydrogenated_1991}, and the resulting materials are denoted as $a$-Si:H. 

The properties of $a$-Si:H have long been rationalized in terms of the underlying structure \cite{elliott_structure_1989}. Adding hydrogen formally disrupts the (idealized) fourfold-connected continuous random network of Si atoms \cite{Wooten-85-04}, and leads to the passivation of defect states such that $a$-Si:H structure models show a clean gap in the electronic density of states (DOS) \cite{legesse_revisiting_2013, nepal_new_2025}. Hydrogen enhances the stability and carrier mobility in $a$-Si:H \cite{street_hydrogenated_1991}, as well as reduces electron-hole recombinations at defect sites, providing higher conversion efficiency in p--n solar cells \cite{shah_thin-film_2012}. A challenge in exploring the effect of hydrogenating silicon has been the cost of first-principles (ab initio) simulation methods and the need for very short molecular-dynamics (MD) timesteps to describe the motion of the light hydrogen atoms. Thus, many open questions still remain around $a$-Si:H, such as its degradation under light, known as the Staebler--Wronski effect \cite{staebler_reversible_1977}, or the nature of ``protocrystalline'' solar cells \cite{ishikawa_flexible_2006}. 

Atomistic modeling of $a$-Si and, to a lesser extent, $a$-Si:H has been accelerated by machine-learned interatomic potential (MLIP) models \cite{caro_machine_2023}. An important milestone in this regard has been a general-purpose Gaussian approximation potential (GAP) for elemental silicon \cite{Bartok-18-12}, subsequently denoted ``Si-GAP-18'': this model has been applied to the study of ambient- and high-pressure disordered forms of silicon \cite{Deringer-18-06, Bernstein-19, Deringer-21-01}. It was later shown how Si-GAP-18 could be distilled into a faster ``student'' model \cite{morrow_indirect_2022}, and the latter has been applied to study defects in a million-atom simulation of $a$-Si \cite{morrow_understanding_2024} as well as the presence of local structural order in the amorphous phase \cite{rosset_signatures_2025}. There was also a report of a GAP model for $a$-Si:H \cite{unruh_gaussian_2022}, to which we refer as ``SiH-GAP-22'' in the following, which was applied to modeling crystalline-/$a$-Si:H interfaces \cite{diggs_hydrogen-induced_2023}. However, the latter model was deliberately focused on the disordered phases, and a full MLIP-driven exploration of the binary Si--H system has not yet been reported to the best of our knowledge.

Drawing inspiration from the unified description of the Si--O system achieved in a previous study \cite{Erhard-24-03} with the atomic cluster expansion (ACE) framework \cite{drautz2019atomic, lysogorskiy2021performant, bochkarev2022efficient}, we now set out to explore the Si--H system. Indeed, bespoke ACE-based MLIPs have reached highly competitive simulation length and time scales and have provided insights into complex structural transitions \cite{ Erhard-25-03, zhou_full-cycle_2025}; a recent benchmark across a range of MLIP fitting frameworks showed that nonlinear ACE models enable fast inference while maintaining good accuracy \cite{leimeroth_machine-learning_2025}.

In the present work, we introduce an MLIP model for the binary Si--H system based on the ACE framework. 
We describe the development of our training dataset for Si--H through an active-learning cycle, the careful validation of the model against DFT reference data, and comparison with experimental observables where applicable. We evaluate the potential's performance across bulk and surface benchmarks and illustrate its usefulness by studying the mechanical properties of $a$-Si:H.

\begin{figure*}
    \centering
    \includegraphics{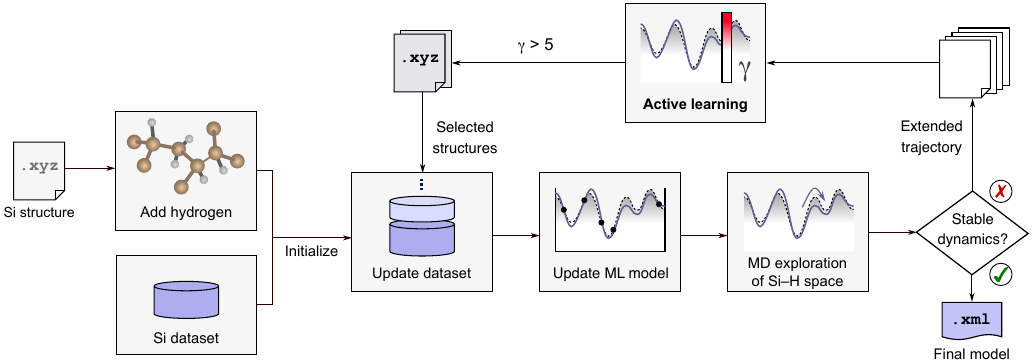}
    \caption{Simplified overview of the approach for MLIP training dataset construction, including iterative exploration and ACE model fitting, adopted in the present study.}
    \label{fig:overview}
\end{figure*}

\section{Methodology}

\subsection{A training dataset for the Si--H system}

Figure \ref{fig:overview} presents the overall approach for the generation of the reference dataset and the iterative training of the potential using an active-learning (AL) workflow.
AL aims to iteratively expand a potential's training dataset in regions of the configurational space where its predictions are uncertain. A range of AL strategies have been proposed \cite{smith_less_2018, Kulichenko-23-03, lysogorskiy2023active,  vanderOord-23-09, Zaverkin-24-04, Perego-24-12}, typically relying on uncertainty metrics that assess the potential's confidence in its own predictions \cite{Novikov-21-01, Grasselli-25}.

As an initial dataset, we took large $a$-Si structures generated in a previous study \cite{rosset_signatures_2025}, to which we added
crystalline Si configurations obtained by systematically scaling and distorting equilibrium cells, and randomly displacing atoms from their equilibrium positions. We also added liquid Si structures of varying densities from separate MD runs at 5,000 and 9,000 K, respectively. 

To supplement this dataset with Si--H configurations, we added SiH$_4$, Si$_2$H$_6$, and Si$_3$H$_8$ molecules as well as SiH$_2$ and SiH$_3$ radicals. Using a simple script for adding hydrogen, we generated additional diamond-type, liquid, and amorphous structures, each hydrogenated with up to 20 at.-\% H. The hydrogenation script used rules informed by literature \cite{Acco-96-02, danesh_hydrogen_2004, shkrebtii_dynamics_2010, jarolimek_first-principles_2009,Powell-89-12}: adding H to dangling Si bonds, creating voids by removing Si atoms and decorating the new coordination defects, or placing H$_2$ molecules in voids. Some H atoms were also introduced as interstitial or substitutional defects. We note that hydrogen can adopt a wealth of conformations in the silicon matrix \cite{ching_electronic_1980}, and that the script only provided a starting point for hydrogenation.

Prior to fitting, we applied filters to our initial dataset and to all subsequent additions made via AL. We removed all structures with positive DFT energies relative to the energy of isolated Si and H atoms, with a maximum force magnitude of $> 50$ eV/Å, with Si--Si distances of $< 1.6$ Å, or with Si--H distances of $< 1$ Å. Details of the DFT computations are described in Appendix A.
We then fit an initial MLIP model to this ``Iter-0'' dataset, and used it to extend the breadth of the configurational space covered by the dataset.

\begin{figure*}
    \centering
    \includegraphics[width=0.8\linewidth]{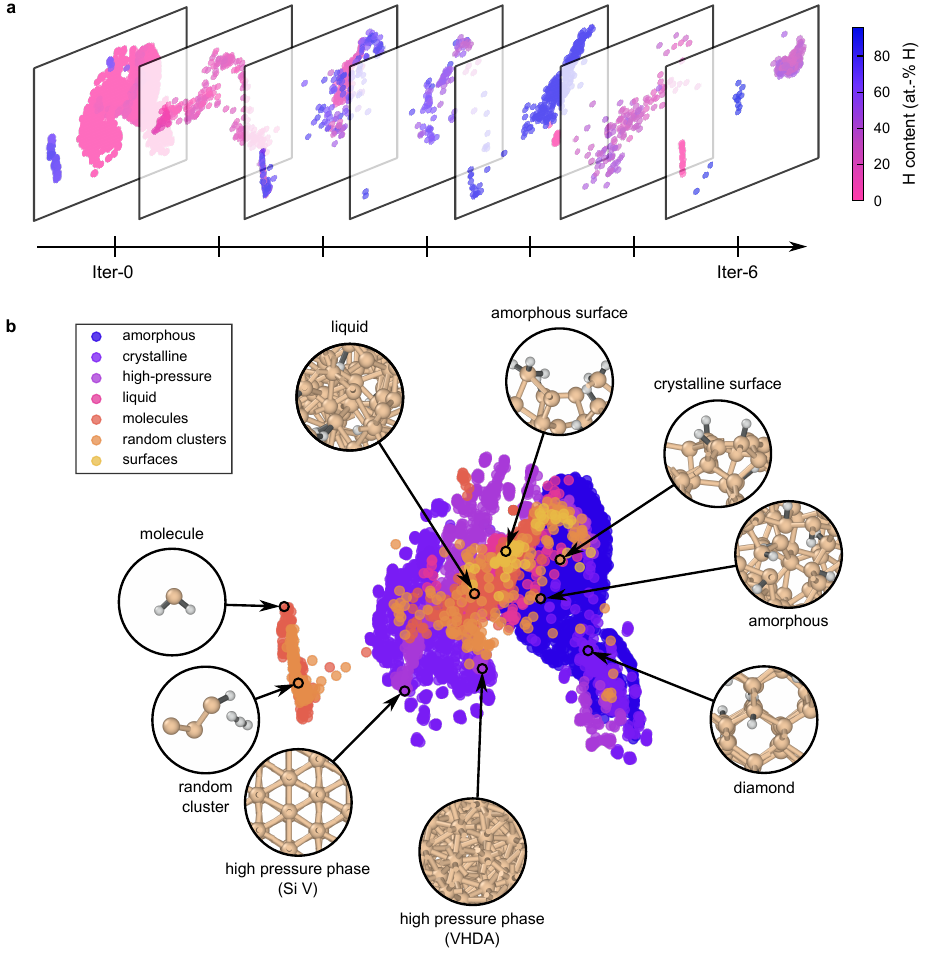}
    \caption{
    The SiH-25 dataset.
    (a) Training dataset evolution over active learning iterations, color-coded by hydrogen content, visualized using UMAP \cite{mcinnes_umap_2018} and SOAP descriptors \cite{Bartok-13-05}.
    (b) Projected map of the full dataset, where each point represents a structure in the training set and is color-coded by structure type. Representative structures of each configurational type, visualized using OVITO \cite{ovito}, are shown as insets.}
    \label{fig:dataset}
\end{figure*}

To choose which structures to add to the dataset with AL, we used the extrapolation grade, $\gamma$ \cite{lysogorskiy2023active}, based on the D-optimality criterion \cite{podryabinkin_active_2017} as a measure of the interpolation ($\gamma<1$) or extrapolation ($\gamma>1$) behavior of the model on an unseen configuration, relative to the subset of its training dataset that spans the widest space. 
By evaluating $\gamma$ over MD simulations, we selected all structures encountered during the simulation that had $\gamma>5$, i.e., a value placing the structure firmly outside the interpolation regime, and then performed Farthest Point Sampling on this set to determine the subset of structures that would provide the widest sampling of the space. 

Once the potential showed stable dynamics for amorphous and crystalline configurations, we simulated surface structures by cleaving and relaxing bulk amorphous structures. We also decorated crystalline (diamond-type) Si surfaces with hydrogen, specifically the (100), (111), (221), and (311) surfaces.
We added small random clusters by random structure search and relaxation, as these have been shown to substantially improve model stability in MD \cite{zhou_full-cycle_2025}. We stopped AL when we no longer encountered $\gamma>5$ configurations when performing MD in relevant phase space. The potential performed well across the Si--H space and would no longer be expected to benefit substantially from additional training structures.

The complete dataset composition, by configuration type and AL iteration, is detailed in Table~\ref{tab:dataset-comp} and visualized in Fig.~\ref{fig:dataset}. As a metric of comparison, the Si-GAP-18 training dataset contains 2,475 structures and 171,815 atomic environments \cite{Bartok-18-12}, while the SiH-GAP-22 training dataset contains 390 structures with 60,770 Si and 5,137 H atoms for a total of 65,907 atomic environments.

\begin{table}[]
    \centering
    \caption{Composition of the training dataset for the SiH-ACE-25 potential, detailing the number of structures and number of atomic environments per element for each configuration type.}
    \setlength{\tabcolsep}{3pt} 
    \begin{tabular}{lccccc}
        \hline\hline
        \multirow{2}{*}{Type}& \multirow{2}{*}{Structures} & \multicolumn{3}{c}{Atomic environments} & \\
          &  & \makecell{Si} & \makecell{H} & \makecell{Total} \\
        \hline
        Amorphous      & 2,030 & 265,453 & 14,233 & 279,686  \\
        Liquid         & 697 & 56,515 & 12,450 & 68,965  \\
        Crystalline    & 1,289 & 51,296 & 2,995 & 54,291  \\
        High pressure  & 717 & 16,776 & -- & 16,776  \\
        Molecules       & 438 & 997 & 4,487 & 5,484  \\
        Surfaces        & 54 & 2,812 & 2,550 & 5,362  \\
        Random clusters & 382 & 808 & 1650 & 2,458 \\
        \hline
        Total & \bf{5,607} & \bf{394,657} & \bf{38,365} &\bf{433,022} &\\
        \hline\hline
    \end{tabular}
    \label{tab:dataset-comp}
\end{table}

We set aside 20\% of the entire set of structures generated by the protocol as a validation set, splitting by configuration type to enforce proportional representation.
We also defined two further test sets: 10\% of the training set of Si-GAP-18, to which we refer as ``Si-GAP-18-set'', and the totality of the training set of SiH-GAP-22 (``SiH-GAP-22-set''). We evaluate our model on these training sets as they are more diverse and comprehensive than the test sets defined in the original publications \cite{Bartok-18-12, unruh_gaussian_2022}: these test sets only contain selected configurations, rather than splits from the respective full training sets.

\subsection{ACE model parameterization}

With the training dataset in place, we turn to the parameterization of the potential.
A particularity of our training dataset is that it contains $10 \times$ more Si environments than H environments (Table \ref{tab:dataset-comp}), and that these Si environments are very diverse, as illustrated in Fig.~\ref{fig:dataset}. This focus on Si is a reflection of our interest in hydrogenated silicon phases, with hydrogen content below 30 at.-\% H. To best describe the Si--H system, we opt for a custom potential shape that contains different numbers of basis functions for each element, unlike some other multi-element ACE potentials reported earlier \cite{nicholas_structure_2025, zhou_full-cycle_2025}. We compare the numerical predictions of these custom, ``asymmetric'' models to ``symmetric'' models of the same total size in Appendix B, and show that the asymmetric models perform marginally better.

\begin{figure}
    \centering
    \includegraphics[]{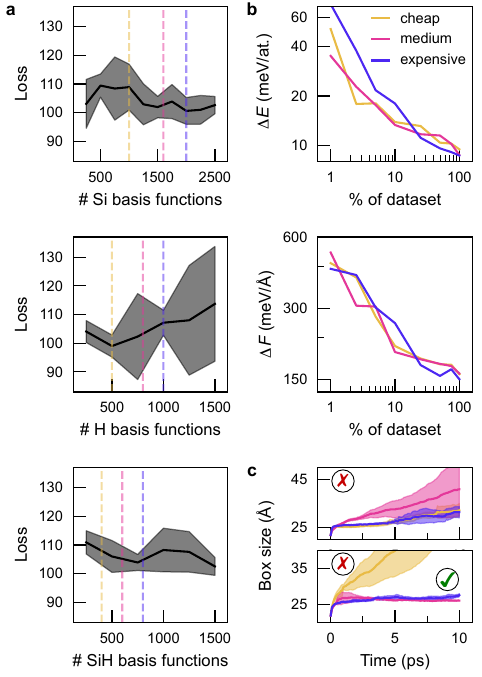}
    \caption{(a) Combined energy and force loss against number of elemental basis functions for Si (top), H (middle) and SiH (bottom), where we plot the average of five randomly initiated models and the standard deviation in gray. Colored vertical dashed line indicate the basis sizes for three models of increasing cost. (b) Energy and force RMSE as a function of dataset size for three models of increasing cost, where we systematically pick the best of 5 randomly initiated models, presented as log--log plots. (c) Evolution of box size against simulation time for three models of increasing cost, plotting the average and the standard deviation of five repeats for the ``symmetric'' models (top) and ``asymmetric'' models (bottom).}
    \label{fig:ablations}
\end{figure}

To inform our choice of model parameterization, we systematically vary the number of basis functions per block\footnote{Basis functions are organized by species blocks. An elemental model will have a singular block, while an A--B binary model will have 4 blocks: A--A, B--B, A--B and B--A.}, while keeping the other three blocks fixed at 1600, 800 and 600 for Si, H and SiH respectively, and train five randomly initialized models for each parameterization. As our interest lies in the overall trends, we only use a 20\% split of our total training dataset to fit these models. For each basis size, we evaluate a combined energy and force loss function on the SiH-GAP-22-set [Fig.~\ref{fig:ablations}(a)]. 

For both the Si and SiH elemental blocks, increasing the number of functions and hence the complexity of the model does not offer a substantial decrease in the loss. Conversely, increasing the number of H basis functions first improves, then notably {\em worsens} the loss -- this could reflect an overfitting scenario related to the dataset's heterogeneity. Figure \ref{fig:ablations}(a) warns against too many H basis functions, but does not otherwise indicate a preferential basis size for the Si and SiH blocks.

To further guide our choice, we fit three potentials of increasing computational cost: a ``cheap'' model with 2,300 total functions, a ``medium'' model with 3,600 total functions, and an ``expensive'' model with 4,600 total functions, where the elemental basis block sizes are shown as vertical lines in Fig.~\ref{fig:ablations}(a). 
As a metric of cost comparison, the ``cheap'', ``medium'' and ``expensive'' potentials can run 1.825, 1.284 and 1.084 ns/day, respectively, for a system of 1,000 atoms and a timestep of 0.1 fs on an NVIDIA A100 GPU card.

We then compare the individual energy and force predictions from these three models on the SiH-GAP-22-set, in Fig.~\ref{fig:ablations}(b), and ensure that the results are converged for the size of the training dataset, such that increasing the dataset size does not substantially improve the energy or force RMSE. All three models perform comparably in terms of energy and force metrics, achieving energy errors under 10 meV/at.\ and force errors around 150 meV/Å.

We move beyond numerical testing with a domain-specific test on a difficult benchmark: a 10 ps simulation in the NpT ensemble of a liquid Si--H system of 1,000 atoms with 50 at.-\% H at 2,000 K. This constitutes a challenging test as hydrogen is highly mobile and diffuses quickly at this elevated temperature, leading to short distances and high forces.
We assess the stability of the MD simulations by plotting the average box size as a function of simulation time, with five repeats of the protocol for each model, comparing our ``asymmetric'' models to their ``symmetric'' counterparts [Fig.~\ref{fig:ablations}(c)]. 

Despite their good numerical accuracy (Appendix B), all three symmetric models are unstable in MD: the simulation box quickly expands unreasonably (``explodes'') for the ``medium'' model, while the box expansion has a delayed onset of around 5 ps for the ``cheap'' and ``expensive'' models. Similarly, the ``cheap'' asymmetric potential is too simple to stabilize the liquid Si--H mixture at 2,000 K, and the simulation box explodes. However, the ``medium'' and ``expensive'' asymmetric potentials do capture the correct dynamics of this challenging test scenario. This justifies the choice of a custom potential shape: for the same overall model size, the ``medium'' and ``expensive'' asymmetric models are stable while their symmetric counterparts are not.

As the ``medium'' and ``expensive'' potentials perform comparably well both numerically and physically, we select the ``medium'' potential shape for its reduced complexity and cost. With this custom ``asymmetric'' shape, a final model was fitted with optimized hyperparameters (Appendix A), hereafter denoted as ``SiH-ACE-25''. 

\subsection{The SiH-ACE-25 model}

Table~\ref{tab:test-err} details the SiH-ACE-25 model's root-mean-square error (RMSE) values for energy and force predictions across all three validation sets considered. The errors on our validation set have been separated out by structure type to highlight which of the configuration types are more challenging to describe. 

\begin{table}[]
    \centering
    \caption{Predictions of the SiH-ACE-25 potential across three datasets: a validation set obtained as 20\% of our AL-generated dataset; a 10\% split of the Si-GAP-18 training set; and the SiH-GAP-22 training set. We detail validation-set composition by structure and atomic environments, and provide energy and force-component RMSE values.}
    \setlength{\tabcolsep}{6pt}
    \begin{tabular}{lcccc}
        \hline\hline
        \multirow{2}{*}{Type}  & \multirow{2}{*}{\makecell{$\Delta E$ \\ (meV/at.)}} & \multicolumn{3}{c}{$\Delta F$ (eV/Å)} \\
        \cline{3-5}
        & & Si & H & All \\
        \hline 
        Amorphous     & 11.5 & 0.12 & 0.23 & 0.13 \\
        Liquid         & 19.8 & 0.22 & 0.44 & 0.28  \\
        Crystalline    & 11.9 & 0.11 & 0.21 & 0.12  \\
        High pressure\footnote[1]{Si structures only}  & 14.1 & 0.22 & -- & 0.22 \\
        Molecules      & 79.4 & 0.82 & 0.44 & 0.53  \\
        Surfaces        & 24.4 & 0.24 & 0.18 & 0.22 \\
        Random clusters & 147.0 & 0.80 & 0.61 & 0.67  \\
        Total  & \bf{46.3} & \bf{0.15} & \bf{0.36} & \bf{0.18}  \\
        \hline
        Si-GAP-18-set & 24.5 & 0.13 & -- & 0.13 \\
        \hline
        SiH-GAP-22-set & 8.7 & 0.15 & 0.30 & 0.16  \\
        \hline\hline
    \end{tabular}
    \label{tab:test-err}
\end{table}

From our validation dataset, the more disordered configurations generally incur higher errors, with the random clusters (along with molecules) having the largest errors likely as they contain larger forces on average than other configurations. Furthermore, these configurations account for less than 2$\%$ of the total atomic environments seen in training (cf.\ Table~\ref{tab:dataset-comp}); more training data might reduce these errors.
The hydrogen environments are more challenging for the potential, as seen by the higher force errors on H than Si, by about a factor of 2 for bulk configurations. This stems from the unbalanced dataset that contains far fewer H atomic environments. Overall, the model performs well across the dataset, particularly on bulk amorphous and crystalline configurations.

We further test the model's performance on the Si-GAP-18-set and SiH-GAP-22-set validation sets. The Si-GAP-18-set contains 229 structures and 52,630 atomic environments, while the SiH-GAP-22-set contains 358 structures and 52,630 atomic environments, of which only 3,698 are H environments. The potential shows very good performance on both sets, with much lower average errors than on our validation dataset. This is likely due to the fact that the configurational space sampled in these datasets is more limited, such that our validation set is a more challenging numerical benchmark for the potential.

\section{Validation}
\subsection{Crystalline-phase properties}

We first evaluate the performance of our model on a range of properties of the crystalline diamond-type modification of silicon, \textbf{dia}-Si (Table~\ref{tab:cSi}). We include formation energies for hydrogen defects in the crystalline network, reporting values for the two most common defect types: interstitial H atoms and H$_2$ molecules \cite{van_de_walle_energies_1994}. We note that these configurations have not been explicitly included in the training set.

\begingroup
\setlength{\tabcolsep}{7pt}
\begin{table}[t]
    \centering
    \caption{Properties of diamond-type silicon, comparing SiH-ACE-25 predictions to SCAN and literature values for the lattice constant ($a_0$), the atomization energy ($E_0$) and the bulk modulus ($B$) of diamond-type silicon, and the formation energies of a silicon vacancy (V$_\mathrm{{Si}}$), a hydrogen interstitial ($\mathrm{{H}}_\mathrm{{i}}$), and a H$_2$ interstitial (${\mathrm{{H}}_2}_\mathrm{{i}}$).}
    \begin{tabular}{lccc}
        \hline\hline
          & \multicolumn{2}{c}{This work} & \\
          \cline{2-3}
          & SCAN & SiH-ACE-25   & Expt. (Lit.) \\
        \hline
         $a_0$ (Å) & 5.43 & 5.42  & 5.43 \cite{hubbard_silicon_1975} \\
         $E_0$ (eV/at.) & --5.86 &  --5.86 & -- \\
         $B$  (GPa)  & 94.17 & 98.08 & 99.0 \cite{mcskimin_measurement_1953} \\
        \hline\hline
          & \multicolumn{2}{c}{This work} & \\
          \cline{2-3}
          & SCAN & SiH-ACE-25 & DFT (Lit.) \\
        \hline
        V$_\mathrm{{Si}}$ (eV)  &  3.94 & 3.41 & 3.17 \cite{probert_improving_2003}\\
        $\mathrm{{H}}_\mathrm{{i}}$ (eV)  & 1.56 & 0.55 & 1.04 \cite{morris_hydrogensilicon_2008}\\
        ${\mathrm{{H}}_2}_\mathrm{{i}}$ (eV/at.) & 0.66 & 0.23  & 0.80 \cite{van_de_walle_energetics_1999}\\
        \hline\hline
    \end{tabular}
    \label{tab:cSi}
\end{table}
\endgroup 

The SiH-ACE-25 model provides excellent agreement with experiment for the lattice parameter, $a_{0}$, of \textbf{dia}-Si. It also provides excellent agreement with DFT for the atomization energy (relative to an isolated Si atom), $E_0$. The potential's prediction of the bulk modulus, $B$, closely matches experiment. We note that a benchmark study of estimated bulk moduli of \textbf{dia}-Si found a variation of about 10 GPa across different DFT functionals \cite{barhoumi_elastic_2021}.

We next turn to the energies of various defects in \textbf{dia}-Si. We focus on the relaxed structures and the relative ordering of defect energies, rather than on the absolute values as these vary strongly with DFT functionals and parameterizations \cite{Corsetti-11-07}. 
The relaxation around silicon vacancy (V$_\mathrm{{Si}}$) defects is a well documented, challenging test that requires large structural models and well-converged parameters, or will otherwise lead to incorrect outward relaxation behavior \cite{probert_improving_2003}.
SiH-ACE-25 correctly predicts the Si atoms to relax inward around the V$_\mathrm{{Si}}$ site, with an inward movement of magnitude 0.28 Å, compared to a SCAN-DFT prediction of a magnitude of 0.16 Å.

Interstitial H atoms adopt a bond-centered configuration, wherein the $\mathrm{{H}}_\mathrm{{i}}$ defect sits at the center of a nearest-neighbor Si--Si bond, causing a substantial outward relaxation \cite{DeLeo-88-10}. Our model reproduces the correct relaxed symmetry, stretching the bond from 2.35 to 3.19 Å. This compares to a relaxed bond length of 3.20 Å from DFT and 3.25 Å from literature \cite{DeLeo-88-10}. However, the energy prediction from SiH-ACE-25 is substantially lower than those from both our SCAN computations and literature.

Finally, we assess the model's predictions of an interstitial H$_2$ molecule in the \textbf{dia}-Si structure. The ${\mathrm{{H}}_2}_\mathrm{{i}}$ energies quoted in Table \ref{tab:cSi} are given per H atom in an interstitial H$_2$ molecule in the bulk, relative to that atom's energy in a H$_2$ molecule in a vacuum. The SiH-ACE-25 potential correctly predicts that an H atom in an H$_2$ interstitial defect is more stable than it would be as a singular $\mathrm{{H}}_\mathrm{{i}}$ defect \cite{van_de_walle_energies_1994, van_de_walle_theory_1989}. The potential relaxes the H$_2$ molecule bond length to 0.774 Å, compared to 0.766 Å for DFT.

\subsection{Amorphous structures}

We use simulated melt-quenching to generate a range of $a$-Si:H configurations with a total of 1,000 atoms each, and hydrogen content ranging from 0 to 50 at.-\% H in increments of 5 at.-\% H, with ten repeats at each concentration for better statistics. Details of the protocol are outlined in Appendix A. We then study the structural properties of these configurations.

\begin{figure}
    \centering
    \includegraphics{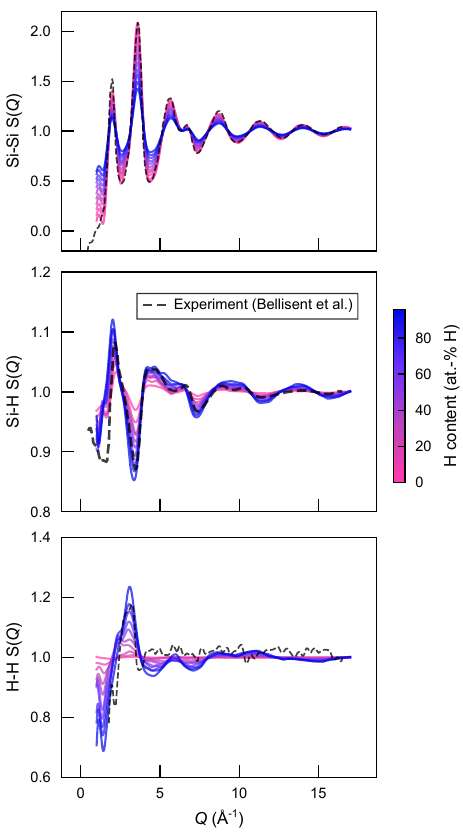}
    \caption{Average partial structure factors of annealed $a$-Si:H structures, showing Si--Si (top), Si--H (middle), and H--H (bottom) contributions. Structure factors are averaged over ten repeats at each concentration. Experimental measurements from Ref.~\citenum{bellisent_structure_1989} for a sample with $\sim$16 at.-\% H are plotted as dashed black lines, normalized for direct comparison.}
    \label{fig:amorph-sq}
\end{figure}

We first plot the partial structure factors in Fig.~\ref{fig:amorph-sq}(a), alongside experimental neutron-scattering data measured for $a$-Si:H samples of 16 at.-\% H made by sputtering (taken from Ref.~\citenum{bellisent_structure_1989}).
Hydrogen immediately affects the connectivity of the silicon backbone network, even in the medium regime of 10 at.-\% H: it disrupts the network by bonding to Si atoms that would otherwise be bonded to other Si atoms, reflected by a broadening of the partial structure factor and decrease of all peak heights. Further increase of the amount of hydrogen in the amorphous matrix further diffuses these peaks. 
Both the partial structure factors for Si--H and H--H sharpen as the hydrogen content in the amorphous matrix increases. Absent at low hydrogen content, a clear signature of the H--H partial structure factor emerges at high hydrogen content, which reflects local ordering of the H atoms.

Comparing to the experimental sample with $\sim$16 at.-\% H, the melt-quenched models match the experimental data for Si--Si and Si--H structure factors well, reflecting the quality in prediction of the underlying amorphous matrix. Aside from the first sharp peak, the experimental H--H signal is quite noisy from strong incoherent scattering \cite{bellisent_structure_1989, elliott_structure_1989, wright_neutron_2007}.
While Ref.~\citenum{wright_neutron_2007} constitutes a more recent study of the neutron-scattering data for hydrogenated and deuterated $a$-Si, it only provides a noise- and background-corrected pattern for the deuterated sample, so we opt to compare to Ref.~\citenum{bellisent_structure_1989} instead. Further experimental measurements of the structure factor of $a$-Si:H across preparation methods and a range of hydrogen content would substantially strengthen the comparison to experiment and validation of computational models.

\begin{figure}
    \centering
    \includegraphics[]{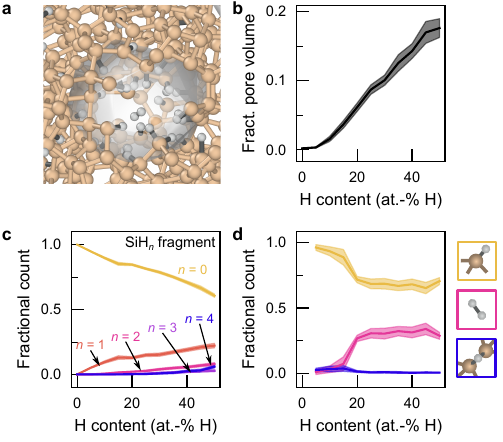}
    \caption{(a) Structure visualization of a pore in an $a$-Si:H sample with 10 at.-\% H, showing the pore edges shaded in gray using the OVITO alpha-shape mesh modifier \cite{ovito, stukowski_computational_2014}. (b) Fractional pore volume in the $a$-Si:H structures as a function of hydrogen content. (c) Count of the occurrence of SiH$_{n}$ fragments in the $a$-Si:H structures relative to the total number of Si atoms, for $n$ taking values from 0 to 4, as a function of hydrogen content in the structures. (d) Count of the occurrence of three different H fragments relative to the total number of H atoms as a function of hydrogen content in the structures, showing the occurrence of H atoms bonded to one Si atom (top, yellow), H atoms bonded to one H atom to form a H$_2$ molecule (middle, pink), and bond-centered H atoms, bonded to two Si atoms (bottom, blue). Visualizations are added to illustrate the structural motifs. We report averaged results over ten repeats at each concentration; shading indicates the standard deviations across repeats.}
    \label{fig:amorph-coord}
\end{figure}

We found that the real-space structural characterization of these $a$-Si:H samples, including silicon coordination, bond angle and ring distributions, is largely unchanged by increasing hydrogen content\footnote{Structural analysis supporting this statement is available at \url{https://github.com/lamr18/SiH-ACE-25}.} -- instead, we turn to a detailed picture of the evolution of atomic environments in $a$-Si:H with increasing hydrogen content in Fig.~\ref{fig:amorph-coord}. Pores in pure $a$-Si are artefacts of the deposition method \cite{Moss-69-11}, and hence are not expected in a simulated quenched structure \cite{Lewis-22-03}. In contrast, for $a$-Si:H, as the hydrogen content increases, pores do appear and grow to account for a substantial fractional volume of the box. Here, we calculate the pore volume by placing a fine 3D mesh on the structure and evaluating the number of empty grid elements, using a script developed in the context of Ref.~\citenum{Liu-25-09}, and present results in Fig.~\ref{fig:amorph-coord}(a). Open and closed porosity are difficult to measure experimentally, but the fractional pore volume in the low-H regime is consistent with experimental observations of pore fractions around 1 to 3\% for samples of H content around 3--8 at.-\% H \cite{mahan_structural_2001}. As a metric of comparison, our 5 and 10 at.-\% H models display average pore fractions of 0.3\% and 1.6\%, respectively. Pore size and pore volume fraction vary with hydrogen content, but also with other parameters of film preparation, such as deposition rate or annealing time \cite{mahan_structural_2001}.

We analyze the local motifs present in the amorphous structures as the hydrogen content is varied, first looking at the occurrence of SiH$_n$ fragments in Fig.~\ref{fig:amorph-coord}(c). 
The preferred structural motif is the monohydride, SiH, fragment, in which a silicon atom is bonded to a singular hydrogen atom (in addition to other silicon atoms). This corresponds to a dangling bond in the Si network that has been passivated by a hydrogen atom. The abundance of this fragment type steadily grows as the hydrogen content increases. As hydrogen can be incorporated into amorphous silicon in excess of the dangling-bond population, the monohydride fragments likely occur in a ``broken-bond'' model wherein a Si--Si bond is broken to accommodate for two monohydride environments \cite{ching_electronic_1980}.

The dihydride (SiH$_2$) fragment is also present, albeit in a much lower proportion, which is consistent with previous reports of --SiH$_2$ on the edges of large voids \cite{melskens_migration_2017}. The fractional count of SiH$_2$ motifs grows steadily alongside the increase in pore volume.
While very few trihydride (SiH$_3$) fragments appear even at very high hydrogen concentrations, a small amount of SiH$_4$ molecules ($\approx6\%$ of H environments) appear at very high hydrogen concentrations, where they detach from the Si network to sit in large voids.

Turning to the hydrogen environments in the amorphous matrix in Fig.~\ref{fig:amorph-coord}(d), the coordination of hydrogen to a single silicon atom is most prevalent (yellow), regardless of the SiH$_n$ fragment. The ``bridge'' conformation \cite{ching_electronic_1980}, shown in blue, has been experimentally observed in $a$-Si:H as a secondary motif \cite{ching_electronic_1980} which agrees with the low proportion of this conformation shown in our quenched structures.

As the hydrogen content increases, H$_2$ molecules form in large voids. Indeed, once an $a$-Si:H structure has been hydrogenated past its solubility limit, excess hydrogen will form H$_2$ molecules \cite{Acco-96-02}. This has been observed experimentally across preparation methods \cite{graebner_solid_1984, carlos_molecular_1982, lohneysen_direct_1984}. While the relative fractional content is not straightforward to resolve experimentally, H$_2$ has been estimated to account for about 13\% of the hydrogen content in films of 7.5 at.-\% H \cite{boyce_orientational_1985}. However, the solubility limit of hydrogen in the disordered silicon matrix varies greatly depending on the preparation method: Ref.~\citenum{Acco-96-02} predicts a solubility limit of 4 at.-\% H for ion-implanted samples, while samples prepared by chemical vapor deposition in Ref.~\citenum{danesh_hydrogen_2004} had a solubility of 16 at.-\% H.

\subsection{Liquid structures}

\begin{figure}
    \centering
    \includegraphics[]{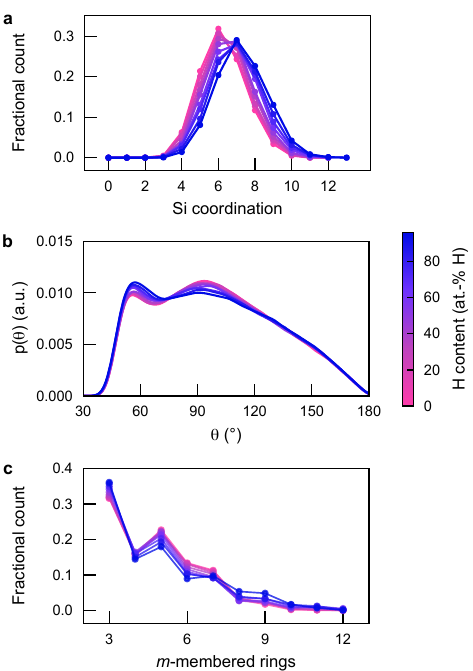}
    \caption{(a) Fractional count of silicon coordination in liquid Si--H mixtures as a function of hydrogen content. (b) Si--Si--Si bond angle distribution. (c) Distribution of \textit{m}-membered shortest-path rings. Distributions are averaged over ten repeats at each concentration.}
    \label{fig:liquid}
\end{figure}

We study the liquid Si--H system by generating a range of liquid structures of 1,000 atoms with increasing H content at a density of 2.58 g cm$^{-3}$, around the estimated density of liquid silicon \cite{dharma-wardana_liquid-liquid_2020, unruh_gaussian_2022}. We equilibrate these configurations at 2,000 K in the NVT ensemble for 10 ps with a timestep of 0.1 fs, running ten repeats for each hydrogen concentration. We plot the results in Fig.~\ref{fig:liquid}, assessing real-space structural characteristics as a function of hydrogen content. 

Figure \ref{fig:liquid}(a) shows the coordination-number distributions for the Si atoms as a function of the hydrogen content in the melt. As the latter increases, the silicon coordination distribution shifts towards higher coordination numbers, from predominantly 6-fold to 7-fold coordinated environments.
The bond-angle distribution trend, shown in Fig.~\ref{fig:liquid}(b), reflects the increase in strained Si environments with an increase in angles of $\approx$~60$^{\circ}$ as the hydrogen content increases. The other dominant peak is at $\approx$~95$^{\circ}$, as determined by peak fitting, and weakens as more strained angles appear.
The count of \textit{m}-membered rings between Si atoms [Fig.~\ref{fig:liquid}(c)] shows the predominance of $m=3$ rings regardless of hydrogen content, and indicates that increasing the hydrogen content reduces the count of intermediate rings ($4\leq m\leq7$) in favor of smaller $m=3$ or larger $m>7$ rings. This further confirms that increasing hydrogen content gives rise to more strained environments. 

This analysis offers insight into the effect of hydrogen on the structure of liquid silicon. One possible explanation is that concentration gradients of hydrogen appear in the melt, resulting in H-rich and H-poor areas. The H-rich areas show larger rings as hydrogen disrupts the Si network due to its low coordination count. The $m=3$ rings would then originate from locally hydrogen-poor regions, where Si atoms have high coordination numbers. The more hydrogen, the greater the ring sizes and pores in the H-rich areas of the network, leading to the local densification of the H-poor areas which see an increase in coordination and $m=3$ ring count.

The Si--H liquid phase is not well characterized in the literature, as most preparation methods for $a$-Si:H rely on molecular precursors and bypass the liquid phase. Liquid silicon adopts complex conformations with several liquid--liquid phase transitions \cite{dharma-wardana_liquid-liquid_2020}, but the effect of hydrogen on these transitions is unknown.

\subsection{Crystalline surfaces}

\begin{figure}
    \centering
    \includegraphics[]{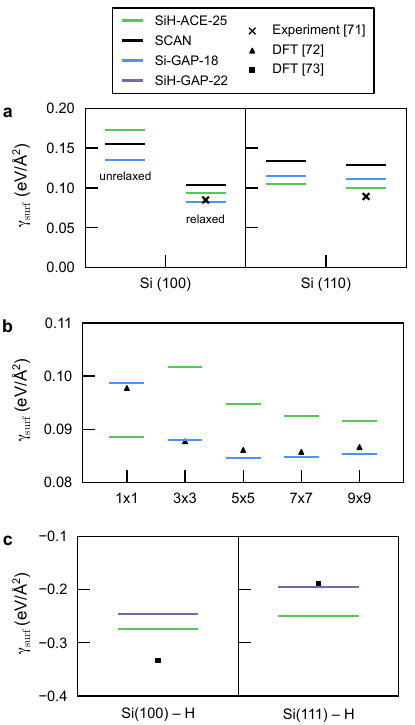}
    \caption{(a) Surface energy, $\gamma_{\text{surf}}$, of the Si(100) and Si(110) surfaces in freshly cleaved and relaxed states, respectively, showing the results for DFT (black), SiH-ACE-25 (green), and Si-GAP-18 (blue) alongside experimental results (black crosses) \cite{eaglesham_equilibrium_1993}. (b) Surface energies for the ideal and reconstructed Si(111) surfaces, for SiH-ACE-25 (green), Si-GAP-18 (blue) and DFT (black triangles) \cite{solares_density_2005}. These Si configurations are unseen for SiH-ACE-25 but seen for Si-GAP-18. (c) Surface energies of the passivated Si(100)--H and Si(111)--H surfaces after relaxation, for SiH-GAP-22 (purple), SiH-ACE-25 (green), and DFT (black squares) \cite{stekolnikov_absolute_2002}.}
    \label{fig:surf-energies}
\end{figure}

We assess the predictions of our model for clean and hydrogen-passivated crystalline (diamond-type) silicon surfaces in Fig.~\ref{fig:surf-energies}. We note that the training dataset includes hydrogenated crystalline surfaces, but no pristine Si ones, which therefore are ``unseen'' in training -- although it is likely that these structures have some surface silicon atoms without any hydrogen in their environment. By comparison, the Si-GAP-18 training dataset does include a large number of diamond-type Si surface configurations \cite{Bartok-18-12}. Adding crystalline silicon surfaces to our training dataset would improve the SiH-ACE-25 model's performance.

We first test our model's performance on unseen clean Si surfaces, by evaluating the surface energies of the cleaved and relaxed Si(100) and Si(110) surfaces, as shown in Fig.~\ref{fig:surf-energies}(a). 
Our potential correctly orders the surfaces by relative stability, but slightly under-stabilizes the Si(100) surface while overstabilizing Si(110) compared to DFT. The model relaxes the (100) surface to its (2$\times$1) reconstruction \cite{ramstad_theoretical_1995}, and simply relaxes the bond lengths of the Si(110) surface, for which there is no consensus on a reconstruction \cite{Waltenburg_1995}. Overall, the potential provides a satisfying match to both DFT and experimental results.

A challenging benchmark for crystalline silicon is the energy of the Si(111) surface and its complex reconstructions \cite{vanderbilt_model_1987}, which we present in Fig.~\ref{fig:surf-energies}(b). The Si(111) reconstructions have been thoroughly studied and the (7$\times$7) reconstruction has been identified as the most stable \cite{zhachuk_strain-induced_2013, lu_relative_2005, vanderbilt_absence_1987}. SiH-ACE-25 incorrectly predicts that the relaxed un-reconstructed (1$\times$1) surface is more stable than the reconstructed surfaces, likely as it has not seen any of these configurations during training. While Si-GAP-18 correctly predicts that the reconstructions are less energetic than the ideal surface, which it has seen in training, even this state-of-the-art potential incorrectly predicts the (5$\times$5) reconstruction as the most stable.
The correct ordering is shown as black squares from DFT calculations in Ref.~\citenum{solares_density_2005}. Reproducing the correct ordering of the energies of the Si(111) surface reconstructions has been a longstanding challenge for both DFT and MLIPs \cite{Bartok-18-12}. The correct ordering was recently achieved with a bespoke neural-network potential targeting the Si(111) surface \cite{hu_atomistic_2021}. Despite this success, the Si(111) reconstructions remain highly non-trivial and complex configurations for any potential.

In Fig.~\ref{fig:surf-energies}(c), we show results for decorating Si(100) and (111) slabs with H atoms and investigating the relaxation behavior. Other, far more complex surface reconstructions with hydrogen have been investigated in the literature \cite{abavare_surface_2014, durr_hydrogen_2013, durr_dissociative_2006, stekolnikov_adatoms_2003}, but Si(100)--H and Si(111)--H are the most commonly occurring hydrogenated surfaces \cite{Waltenburg_1995}. SiH-ACE-25 correctly predicts that the Si(100)--H surface is more stable than the Si(110)--H surface. It adopts the correct dihydride and monohydride reconstructions for the Si(100)-H and Si(110)-H surfaces, respectively. It also provides satisfactory agreement with the DFT computations from Ref.~\citenum{stekolnikov_absolute_2002}.

\subsection{Amorphous surfaces}

\begin{figure}
    \centering
    \includegraphics[]{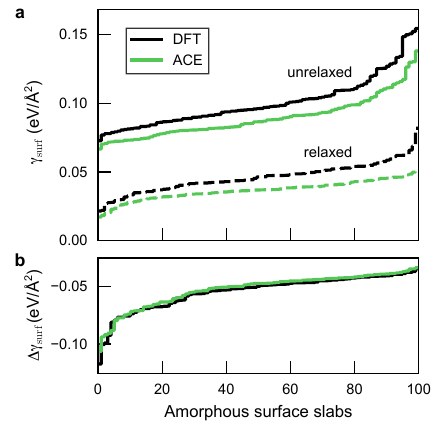}
    \caption{(a) Surface energy, $\gamma_{\text{surf}}$, of 100 amorphous surfaces, freshly cleaved and after relaxation with the SiH-ACE-25 potential, comparing DFT and SiH-ACE-25 predictions. (b) Difference in predicted surface energy before and after relaxation for each model.}
    \label{fig:amorphous-surf-energies}
\end{figure}

Beyond crystalline silicon surfaces, we also assess the model's predictions on amorphous surfaces. We prepare 100 bulk $a$-Si:H structures of varying H content, and cleave them along the $xy$ plane at random distances along the $z$-axis. This results in a set of $a$-Si:H structures of varying size and H content, which we relax with the SiH-ACE-25 potential. We show the distributions of surface energies in Fig.~\ref{fig:amorphous-surf-energies}, alongside the DFT predictions, ordered by stability similar to Ref.~\citenum{Erhard-24-03}.

The distribution of both the unrelaxed and relaxed surface energies predicted by our model is in good agreement with its DFT reference. SiH-ACE-25 faithfully reproduces the distribution of energy changes on relaxation, without large outliers. There is a visible offset in the distribution of absolute surface energies between DFT and our model -- this offset is constant, and we speculate that it could be due to the DFT and SiH-ACE-25 potential's differing treatment of the vacuum and periodic boundary conditions. Inspecting only the energetic effect of relaxation (that is, the surface-energy differences between unrelaxed and relaxed surfaces for individual slabs) confirms that the DFT and SiH-ACE-25 predictions are very similar in this regard [Fig.~\ref{fig:amorphous-surf-energies}(b)].

\begin{figure*}[ht]
    \centering
    \includegraphics[]{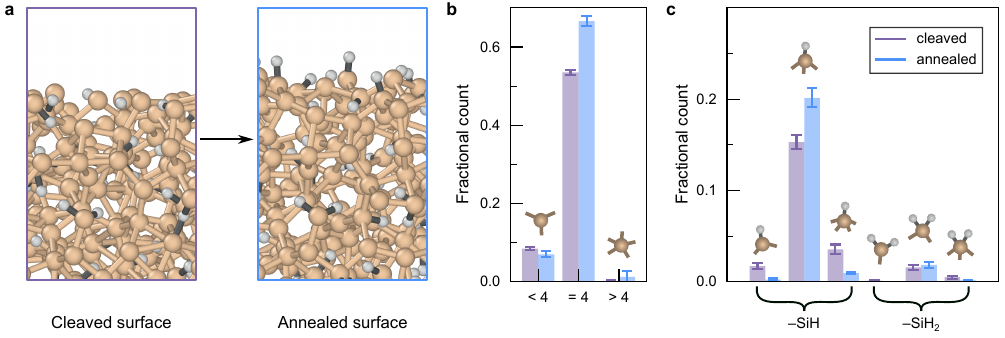}
    \caption{(a) Visualizations of the freshly cleaved $a$-Si:H surface (left) and surface after the annealing treatment (right) with 20 at.-\% H. (b) Histogram of the silicon coordination before and after annealing for different Si environments that are not bonded to any H atoms. (c) Histogram of the silicon coordination before and after annealing for different Si environments bonded to H atoms.}
    \label{fig:amorph-surf-anneal}
\end{figure*}

In Fig.~\ref{fig:amorph-surf-anneal}, we move beyond surface energetics and analyze the structure of a cleaved amorphous surface of 2,500 atoms after an MD annealing treatment. The unrelaxed surface was obtained by cleaving a quenched bulk amorphous structure with 20 at.-\% H along the $xy$ plane. The cleaved structure was then annealed at 500 K for 10 ps in the NVT ensemble, with a timestep of 1 fs. We show structural fragments averaged over ten independent repeats of this protocol. 

As is visible from the renders in Fig.~\ref{fig:amorph-surf-anneal}(a), hydrogen atoms migrate to the surface during the anneal to passivate the dangling Si bonds. The slab also expands along the axis perpendicular to the cut plane, likely to relieve some of the internal stress caused by the surface cleavage.

Figure \ref{fig:amorph-surf-anneal}(b) exclusively considers the bonding environments of silicon atoms that are not bonded to hydrogen. It shows that the freshly cleaved surface contains a large proportion of undercoordinated silicon atoms, or dangling bonds, and far fewer overcoordinated silicon atoms relative to typical well-relaxed amorphous silicon films \cite{morrow_understanding_2024}. Upon annealing, the defect count decreases and the count of ideal, four-fold coordinated, silicon atoms increases to over 65\% of the silicon environments. 

Figure \ref{fig:amorph-surf-anneal}(c) provides a complementary understanding of the local structure after annealing. Within the monohydride fragments, the under- and overcoordinated conformations disappear upon annealing in favor of the typical environment of a silicon atom bonded to three other silicon atoms and a hydrogen atom. The other common fragment is the dihydride fragment, which anneals to a four-fold bonding environment where the silicon atom is bonded to two silicon atoms and two hydrogen atoms.

The low-temperature annealing treatment has lowered the surface energy by 0.072 $\pm$ 0.003 eV/Å$^2$, which is in line with the results from Fig.~\ref{fig:amorphous-surf-energies}.

Concluding the validation thus far, we have shown that the SiH-ACE-25 potential provides good agreement with its DFT reference on amorphous surfaces, and that it can be used to simulate large-scale structural models of surfaces that are inaccessible to direct DFT simulations.

\subsection{Molecules}

We assess the predictions of the SiH-ACE-25 model on radicals and molecular species in Table \ref{tab:molecules}, comparing to our DFT reference. The potential provides excellent agreement with SCAN on optimal bond length and bond angles of the SiH$_4$ molecule, as well as good agreement on these structural metrics for SiH$_2$, SiH$_3$ and Si$_2$H$_6$. This is expected as these configurations are covered by the training dataset. For comparison, SiH-GAP-22 that was not trained on molecules underpredicts the bond lengths across both molecules and radicals, with a relaxed bond length of 1.432 Å for SiH$_4$.

\begingroup
\setlength{\tabcolsep}{3pt}
\begin{table}[t]
    \centering
    \caption{Bond-length ($d$) and bond-angle ($\mathrm{\theta_{H-Si-H}}$) predictions by SCAN and SiH-ACE-25 on molecules and radicals.}
    \begin{tabular}{lccccc}
        \hline\hline
        &  & \multicolumn{2}{c}{$d$ (Å)} & \multicolumn{2}{c}{$\mathrm{\theta_{H-Si-H}}$ ($^{\circ}$)} \\
        & & SCAN & SiH-ACE-25 & SCAN & SiH-ACE-25 \\
        \hline
        SiH$_2$ & & 1.520 &  1.505 & 90.8 & 95.3 \\
        SiH$_3$ & & 1.481 & 1.489  & 111.8 & 107.3 \\
        SiH$_4$ & & 1.479 & 1.481 & 109.5 & 109.5 \\
        \multirow{2}{*}{Si$_2$H$_6$} & Si--Si & 2.329 & 2.339 & \multirow{2}{*}{108.5} & \multirow{2}{*}{109.1} \\
        & Si--H & 1.482 & 1.480 & & \\
        \hline\hline
    \end{tabular}
    \label{tab:molecules}
\end{table}
\endgroup 

\section{Application to $a$-Si:H}

We now illustrate a practical use case for our potential by investigating the fundamental relationship between structure and mechanical properties of $a$-Si:H, a topic that has gathered both experimental and computational interest \cite{bondi_first-principles_2010, jiang_mechanical_2019, kuschnereit_mechanical_1995, johlin_structural_2012}. Indeed, as a solar cell material, $a$-Si:H can face large mechanical loads under weathering conditions, and hence an understanding of the relationship between structure and mechanical properties could help to improve device lifetimes \cite{papargyri_effect_2022}.

\begin{figure*}[ht]
    \centering
    \includegraphics[]{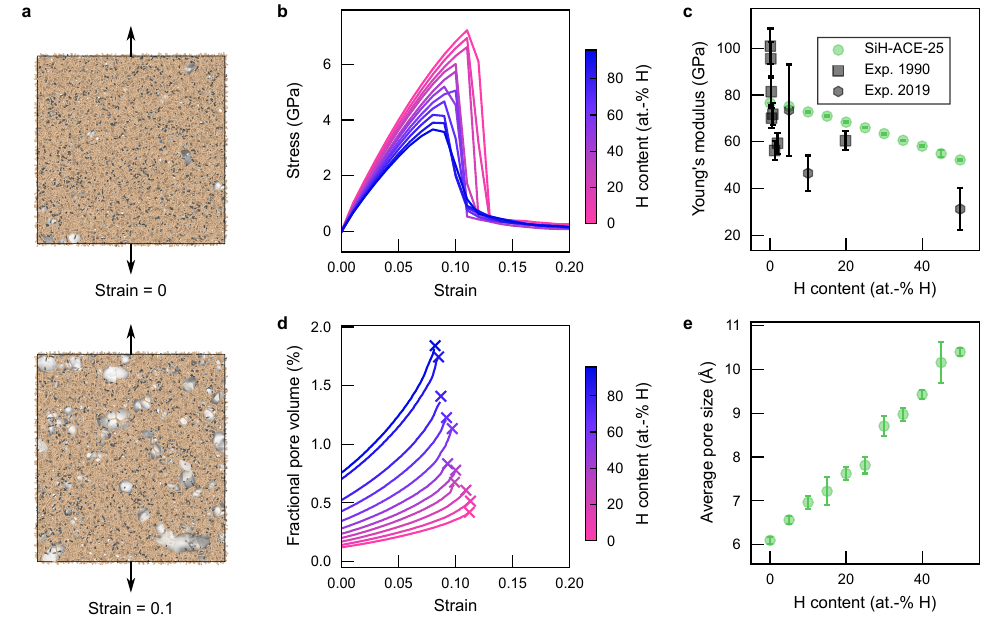}
    \caption{Mechanical and structural properties of $a$-Si:H under uniaxial deformation as a function of hydrogen content. (a) Structure visualization of a 30 at.-\% H $a$-Si:H structure prior to deformation (top) and at a deformation of 10\% strain (bottom), where gray regions show pores using the OVITO alpha-shape mesh modifier \cite{ovito, stukowski_computational_2014}. (b) Stress-strain profiles for $a$-Si:H models of increasing H content, averaged over five repeats at each concentration. (c) Average Young's modulus as a function of H content in the $a$-Si:H network, with error bars showing the standard deviation of each average measurement (green), alongside experimental data from Ref.~\cite{jiang_mechanical_2019} (hexagons) and Ref.~\cite{jiang_mechanical_1990} (squares). (d) Evolution of the average pore volume percentage against strain up to fracture, where fracture is represented as a cross. (e) Average pore size of the deformed material prior to fracture, with error bars showing the standard deviation over five repeats.}
    \label{fig:tensile-testing}
\end{figure*}

We prepare a set of five high-quality $a$-Si configurations of 10,000 atoms by simulated melt-quenching, which we hydrogenate with increasing hydrogen content from 0 to 50 at.-\% H, using the hydrogenation script developed for our dataset generation protocol. We anneal the hydrogenated structures for 10 ps at 500 K in the NpT ensemble.
We then incrementally apply tensile strain along the $z$-axis and perform fixed-volume geometry optimizations, uniaxially deforming the system to a total strain of 0.3 with a strain step of 0.0001. We present the results of our simulated tensile test in Fig.~\ref{fig:tensile-testing}, where we study the effect of H content on the mechanical and structural properties of $a$-Si:H.

Figure \ref{fig:tensile-testing}(a) depicts the uniaxial deformation protocol, showing an $a$-Si:H cell with 30 at.-\% H prior to and after straining to a strain of 0.1, where gray regions highlight the pores in the Si network. 

In Fig.~\ref{fig:tensile-testing}(b--c), we analyze the mechanical properties of our structures, plotting the average stress--strain curve and the Young's modulus over all five repeats. Introducing a small amount of hydrogen into the network immediately decreases the strength of the film, and increasing H content causes the strength to drop further. 
The deformation behavior at low H content shows a longer elastic than plastic deformation window, which agrees with experimental reports on $a$-Si:H samples with 5 at.-\% H \cite{jiang_mechanical_2019}.
The overall shape of the stress--strain curve is unaltered by H content, but with increasing hydrogenation, the film can accommodate less elastic deformation and thus its yield stress decreases considerably.

We estimate the Young's modulus, $E$, of each structure from the gradient of the stress--strain profile in the elastic deformation window, as reported in Fig.~\ref{fig:tensile-testing}(c). We observe a monotonically decreasing trend in elastic strength as the hydrogen content increases, which reproduces other computational results \cite{bondi_first-principles_2010, jiang_mechanical_2019}. 

Mechanical properties of $a$-Si:H films vary greatly with preparation methods, as shown by the experimental data plotted in Fig.~\ref{fig:tensile-testing}(c). The experimental samples in Ref.~\citenum{jiang_mechanical_2019} were made by colloidal synthesis while the samples in Ref.~\citenum{jiang_mechanical_1990} were prepared by radio frequency sputtering. Even for identical preparation conditions, the standard deviations of experimental measurements, shown as error bars in Fig.~\ref{fig:tensile-testing}(c), reflect large variances across samples. As such, our results demonstrate very good agreement with the experimental ranges. 

We assess the evolution of the pore-volume fraction as a function of strain in Fig.~\ref{fig:tensile-testing}(d), plotting the fractional volume up to the point of failure. The pore volume shows constant growth during elastic deformation as the pores are slowly stretched, then a sharp increase during plastic deformation as bonds are broken and new voids appear.

Finally, in Fig.~\ref{fig:tensile-testing}(e), we show the average pore size before rupture as a function of hydrogen content. Experimental findings report a range of pore sizes and shapes, between 1 and 10 Å for Ref.~\citenum{young_nanostructure_2007}, and between 10 to 20 Å for Ref.~\citenum{chabal_molecular_1987}. These wide fluctuations can be attributed to the contributions of both the volume and surface pores, the former usually smaller than the latter, and their relative proportions in the prepared sample \cite{bork_proton_1985}.

Combining the mechanical and structural insights from Fig.~\ref{fig:tensile-testing}, our simulations provide a clear nanoscopic picture of the effect of the hydrogen content in our films: increasing hydrogen content replaces Si--Si bonds with Si--H bonds, disrupting the silicon network and creating voids. Under deformation, a network with high Si--Si connectivity will require higher stress to rupture than a network with low Si-Si connectivity. At low H content, much more energy is required to break bonds and create pores, which will eventually connect and lead to failure. At high H content, pores are already present in the network, and must simply connect to cause failure, so the films yield at lower stress. Increasing H content leads to a porous material, weaker in both the elastic and the plastic deformation regimes. This interpretation is consistent with previous reports of hydrogen weakening the silicon tetrahedral network \cite{bondi_first-principles_2010} and of hydrogen clusters appearing in the silicon matrix \cite{jiang_mechanical_2019}.

Beyond the pores and bonding connectivity, the homogeneity of the distribution of hydrogen within the Si network can also heavily affect the performance of the material under deformation and a heterogeneous distribution can lead to
disparate deformation behavior \cite{bondi_first-principles_2010}. 
A proposed model to understand this inhomogeneity is the formation of H-rich and H-poor domains, with corresponding low- and high-density amorphous networks, as described in Ref.~\citenum{gericke_quantification_2020}. This could explain why the mechanical strength of chemical-vapor-deposited $a$-Si:H films has been reported to peak around 10 at.-\% H, which is also where the films were found to be most dense \cite{kuschnereit_mechanical_1995}.

\section{Conclusion}

We have presented a machine-learned interatomic potential model for the Si--H system, using the nonlinear atomic cluster expansion (ACE) framework to reach inference speeds of $\approx 1$ ns/day for a system of 1,000 atoms on an NVIDIA A100 GPU card. We validated this potential, referred to as SiH-ACE-25, using comprehensive and challenging tests spanning the elemental Si and binary Si--H configurational spaces. We showed that SiH-ACE-25 can be used to create a range of high-quality $a$-Si:H structures of varying hydrogen content, with porosity and coordination environments that coincide with experimental reports. We further assessed the predictions of our potential for bulk crystalline and liquid configurations, as well as crystalline and amorphous surfaces, and made comparisons to experimental data where relevant. We illustrated its use in elucidating structure-property relationships by probing the processes at play during mechanical testing of $a$-Si:H structures.

We expect that the new potential will be helpful for a range of modeling tasks, with particular focus on the $a$-Si:H system. Its computational efficiency promises to unlock simulations at device-sizes and time scales that have never been explored for $a$-Si:H. Aside from the amorphous state, the Si--H liquid phase is not well understood, with very little available literature -- in future work, SiH-ACE-25 could be used to simulate high hydrogen content in the silicon melt, and to study hydrogen dynamics and diffusion in liquid silicon.

Beyond the potential itself and its validation, the development of a large, high-quality dataset of structures across the Si--H configurational space also constitutes a key outcome of the present work. With the rise of ``universal'' \cite{chen_universal_2022, deng_chgnet_2023} or ``foundational'' \cite{batatia_foundation_2024} MLIP models that are trained across a broad range of chemically diverse structures, our dataset could be repurposed (and, if required, relabeled) to create numerical benchmarks to assess the performance of such models specifically on amorphous structures.

\section*{Appendix A: Computational details}

Reference energies and forces were calculated in VASP \cite{kresse_ab_1994, kresse_efficiency_1996}, using the projector augmented-wave formalism \cite{blochl_projector_1994, kresse_efficient_1996}. The SCAN exchange--correlation functional \cite{sun_strongly_2015} was chosen as it has been extensively validated across the configurational space of silicon covering liquid, crystalline and amorphous structures \cite{sun_accurate_2016, remsing_dependence_2017, dharma-wardana_liquid-liquid_2020, Deringer-21-01}. A plane-wave energy cutoff of 1200 eV and a $k$-point spacing of 0.23~Å$^{-1}$ were chosen such that the calculated energies and forces were converged within 1 meV/atom and 10 meV/Å, respectively.

ACE potential fits were carried out using {\tt pacemaker} \cite{lysogorskiy2021performant}. Hyperparameter optimization was performed using Bayesian optimization of a combined energy and force loss function, using the optimization routines implemented in {\tt scikit-optimize} \cite{head_2021_5565057}, through the {\tt XPOT} interface \cite{thomas_du_toit_hyperparameter_2024}. Hyperparameter optimization informed the choice of radial basis parameters and embedding. The full parameterization of SiH-ACE-25 is available at \url{https://github.com/lamr18/SiH-ACE-25}.

We use the SiH-GAP-22 training set as a validation set, relabeling the structures with our chosen DFT parameters. The interface structures of over 400 atoms were too large to label with our DFT parameters, and hence omitted; the reported errors do not include these interfaces.

We perform all MD simulations in LAMMPS \cite{thompson_lammps_2022, Gissinger24}. The $a$-Si:H structures were generated by simulated melt-quenching, using a variable-rate protocol similar to that described in Ref.~\citenum{Bernstein-19}, which reduces computational cost compared to fixed-rate protocols. The structures were quenched in the NpT ensemble from 2,000 to 1,500 K at a rate of 10$^{14}$ K/s, then from 1,500 to 900 K at a rate of 10$^{12}$ K/s, and finally to 500 K at a rate of 10$^{13}$ K/s. The melt-quench protocol was run with an MD timestep of 0.1 fs and repeated ten times at each chosen value of hydrogen content.

For the structural analysis, we used the following cutoff distances, chosen from the tail of the first coordination peak of the RDF of $a$-Si:H: Si--Si=2.85~Å, Si--H=1.8~Å and H--H=1~Å. For the liquid phase, we used: Si--Si=3.1~Å, Si--H=2~Å and H--H=1~Å, similarly chosen from the RDF of the melt.

We use neutron scattering lengths of $b_{\mathrm{Si}}=0.41491\times10^{-14}$~m and $b_{\mathrm{H}}=-0.3739\times10^{-14}$~m, taken from Ref.~\citenum{sears_neutron_1992}.

\section*{Appendix B: Model parameterization} \label{sec:AppB}

To support our choice of a custom ``asymmetric'' model shape, we fit three models of the same total size as the models presented in Fig.~\ref{fig:ablations}(b), but with elemental basis blocks of the same size: that is 575, 900 and 1,150 functions for each of the H, Si, SiH and HSi blocks. Together, these blocks made respective total model sizes of 2,300 functions (``cheap''), 3,600 functions (``medium'') and 4,600 functions (``expensive''). We refer to these models as ``symmetric''.

We compare the numerical performance of these models to our main ``asymmetric'' models in Table \ref{tab:symm-errors}, evaluating the models on the SiH-GAP-22-set. 
The models with an asymmetric potential shape perform marginally better than the symmetric ones for the ``cheap'' and ``medium'' model sizes. This can be rationalized from the greater relative number of Si elemental basis functions in the asymmetric models than in the symmetric models: e.g., for the ``cheap'' models of 2,300 total functions, the symmetric model has 575 Si functions whereas the asymmetric model has 750. As can be expected, a larger number of Si elemental functions will result in better Si force RMSE metrics for Si, but also for the overall force error as there is a majority of Si environments.
The symmetric models show a greater improvement in force errors with increasing number of functions than the asymmetric models, and the ``expensive'' symmetric model performs marginally better than the ``expensive'' asymmetric model.

\begingroup
\setlength{\tabcolsep}{7pt}
\begin{table}[H]
    \centering
    \caption{Force component RMSE values on the SiH-GAP-22-set for three model sizes, comparing a symmetric model parameterization to a custom asymmetric model shape.}
    \begin{tabular}{lcccccc}
        \hline\hline
        Model & \multicolumn{3}{c}{Symmetric} & \multicolumn{3}{c}{Asymmetric} \\
        \hline
         & \multicolumn{3}{c}{$\Delta F_{i}$ (eV/Å)} & \multicolumn{3}{c}{$\Delta F_{i}$ (eV/Å)}\\
         & Si & H & All & Si & H & All \\
        \hline
         cheap   & 0.17 & 0.34 & 0.19 & 0.14 & 0.29 & 0.16 \\
         medium   & 0.16 & 0.29 & 0.17 & 0.14 & 0.30 & 0.16 \\
         expensive   & 0.13 & 0.27 & 0.15 & 0.14 & 0.28 & 0.15 \\
        \hline\hline\\[2mm]
    \end{tabular}
    \label{tab:symm-errors}
\end{table}
\endgroup

\section*{Data availability}
Data supporting the present work, including the SiH-ACE-25 model parameters and training data, are available at \url{https://github.com/lamr18/SiH-ACE-25}. A copy will be deposited in Zenodo upon journal publication.

\begin{acknowledgments}
We thank D.~A.~Drabold, S.~R.~Elliott, and D.~F.~Thomas du Toit for useful discussions, and Y.~Liu for access to his pore analysis code. This work was supported through a UK Research and Innovation Frontier Research grant [grant number EP/X016188/1]. We are grateful for computational support from the UK national high performance computing service, ARCHER2, for which access was obtained via the UKCP consortium and funded by EPSRC grant ref EP/X035891/1. We are also grateful to the UK Materials and Molecular Modelling Hub for computational resources, which is partially funded by EPSRC (EP/T022213/1, EP/W032260/1 and EP/P020194/1).

\end{acknowledgments}

\end{document}